\documentstyle[12pt]{article}
\newcommand{\be}{\begin{equation}}
\newcommand{\ee}{\end{equation}}
\newcommand{\ba}{\begin{eqnarray}}
\newcommand{\ea}{\end{eqnarray}}

\begin{document}
\setlength{\baselineskip}{.7cm}
\renewcommand{\thefootnote}{\fnsymbol{footnote}}
\newcommand{\lp}{\left(}
\newcommand{\rp}{\right)}

\sloppy

\begin{center}
\centering{\bf \Large Large financial crashes}
\end{center}

\begin{center}
\centering{Didier Sornette $^{1-2}$ and Anders Johansen $^3$\\

{\it

$^1$ Department of Earth and Space Science\\ and Institute of Geophysics
and
Planetary Physics\\ University of California, Los Angeles, California
90095\\

$^2$ Laboratoire de Physique de la Mati\`ere Condens\'ee,
CNRS URA 190\\ Universit\'e de Nice-Sophia Antipolis, B.P. 71, Parc
Valrose, 06108 Nice Cedex 2, France\\

$^3$ CATS, Niels Bohr Institute, Blegdamsvej 17, Copenhagen 2100, Denmark}
}
\end{center}
\renewcommand{\thefootnote}{\fnsymbol{footnote}}

{\bf Abstract} \\
We propose that large stock market crashes are analogous to critical points
studied
in statistical physics with log-periodic correction to scaling. We extend
our
previous renormalization group model of stock market prices prior to and
after
crashes [D. Sornette et al., J.Phys.I France 6, 167, 1996]  by including
the first
non-linear correction. This predicts the existence of a log-frequency shift
over time
in the log-periodic oscillations prior to a crash.  This is tested on the
two largest
historical crashes of the century, the october 1929 and october 1987
crashes, by
fitting the stock market index over an interval of 8 years prior to the
crashes. The
good quality of the fits, as well as the consistency of the parameter
values obtained
from the two crashes, promote the theory that crashes have their origin in
the
collective ``crowd'' behavior of many interacting agents.

\vskip 0.5 cm

\pagebreak

Stock markets are fascinating structures with analogies with arguably the
most complex dynamical system found in the Natural Sciences 
\cite{Anderson}, {\it i.e.}, the human mind. Stock market prices weave
patterns on many
different scales, luring traders to believe (maybe correctly) that some
predictability is possible \cite{Prost,Bechu}.
Many attempts have been made to  model stock markets and recently a wealth
of works
from the physical community  has pointed at similarities between stock
markets and
dynamical  out-of-equilibrium systems [4-31].

In this work, we will focus on the most extreme behavior of stock markets,
namely the two largest financial crashes in this century. Our Ariadne's
thread is that
complex systems often reveal more of their structure and organization in
highly
stressed situations than in equilibrium. Hence, our hope is that the study
of these
two large crashes will enable us to extract important new information about
the
dynamics of stock markets. Specifically, we are interested in describing
the stock
market behavior before and after a crash and our problem thus belongs to
the more
general problem of describing the transient behavior preceding a final
equilibrium
state assuming it exists. Our point of view is influenced by the concept of
criticality
developed in statistical physics in  the last 30 years in order to describe
a class
of cooperative phenomena, such  as magnetism and melting, and our
hypothesis
is that
the stock market behaves  as a driven out-of-equilibrium many-body system 
\cite{Anderson,Nicola}

>From the open on wednesday 23 October 1929 to the close of tuesday 29
October 1929,
the New York Stock Exchange lost almost $30 \%$ of its value. An often
quoted origin
of the crash is that traders thought that the bullish trend was due to
continue,
while the efficiency eventually brought back the market  to its
fundamentals. In a
similar fashion, major indexes of market valuation  in the United States
declined by
30 percent or more from the opening on  October 14 1987 to the market close
on
october 19 and in addition all major world markets declined substantially
in the
following month, in contrasts  with the usual modest correlations of
returns across
countries. A lot of work has been carried out to unravel the origin(s) of
the 1987
crash, notably in the properties of trading and the structure of markets;
however, no
clear cause has been singled out \cite{Barro}. Maybe the most quoted
scenario
involves the role of portfolio insurance strategies in amplifying the
descent.  In the
present work, we would like to defend the thesis that these two crashes
have
fundamentally similar origins, which must be found in the collective
organization of
the market traders leading to a regime known as a ``critical''  point.

In a previous paper \cite{Sornette}, we have identified precursory
patterns, as well
as aftershock signatures and characteristic oscillations of relaxation,
associated
with the October 1987 crash up to $2.5$ years in advance. Very similar
results were
obtained  \cite{Feigen} for both the 1929 and 1987 crashes and here it
was pointed out that the log-frequency of the observed log-periodic
oscillations
seems to decrease as the time of crash is approached.

In the present work, we will further substantiate that the concept of
criticality can
by applied to stock market crashes. We generalize our previous work
\cite{Sornette}
and show how the first generic nonlinear correction to the renormalization
group equation previously proposed accounts quantitatively for the behavior
of the
Dow Jones and $S\&P500$ (Standard and Poor) indices up to $8$ years prior
to the
crashes of 1929 and 1987. The conclusion is that the qualitative
observation of 
\cite{Feigen} of the log-frequency shift is rationalized by the
nonlinear correction.

\section{Model and nonlinear generalization}

Using the renormalization group (RG) formalism on the stock market index
$I$ amounts
to assuming that the index at a given time $t$ is related to that at
another time
$t'$ by the transformations \cite{Sornette}
\be
x' = \phi(x)   ,
\label{firstt}
\ee
\be
F(x) = g(x) + \frac{1}{\mu} F\biggl(\phi(x)\biggl),
\label{secondd}
\ee
where $x=t_c-t$.  $t_c$ is the time of global crash and $\phi$ is called
the RG flow map. Here,
$$
F(x)=I(t_c)-I(t)
$$
such that $F=0$ at the critical point and $\mu$ is a constant describing
the scaling of the index evolution upon a rescaling of time (\ref{firstt}).
The function $g(x)$ represents the non-singular part of the function
$F(x)$. We assume as usual \cite{Goldenfeld} that the function $F(x)$ is
continuous and that $\phi(x)$ is differentiable.

As the simplest non-trivial solution beyond the pure power law solution to
(\ref{secondd}), we have proposed \cite{Sornette}
\be
I\lp t \rp  = A+B\lp t_c-t \rp ^{\alpha} \left[ 1+C\cos \lp
\omega \log \lp t_c-t \rp - \phi \rp \right] .
\label{sixthh}
\ee
It includes the first Fourier component of a general log-periodic
correction to a
pure power law behavior of an observable (here, the stock market index)
exhibiting a
singular behavior at the time $t_c$ of the crash, {\it i.e.}, which becomes
scale-invariant at the critical point $t_c$. We have found (\ref{sixthh})
to fit the
S\&P500 data prior to the 1987 crash very well over a period of
approximately two
years before the crash.
We have also tested (\ref{sixthh}) on a few other minor crashes after
1987\footnote{These smaller crashes do not have a very long build-up time
as the 1929
and 1987 crashes and the number of oscillations observed prior to these
crashes are
only one or two. Hence, the need for a log-periodic correction is not as
obvious as
with the two large crashes and we have not included the analysis of these
smaller
crashes here. In analogy with usual critical phenomena, we expect that the
time
interval over which the precursors of a crash are detectable (corresponding
to the
so-called width of the critical region) increases with the size of the
crash. This
rationalizes the strength of the log-periodic corrections for the two large
crashes of
the century.}. Expression (\ref{sixthh}) accounts very well for the two
most
important structures that are visible with the naked eye, {\it i.e.}, the
overall
accelerated increase before the crash as much as two years before it and
its
decoration by large scale oscillations whose linear frequency increases on
the
approach to the crash.

We now proceed to extend these results in order to show that precursors
can be identified over a much longer time interval and that their structure
can be deduced from a general renormalization group approach that we now
present. In this goal, we notice that the solution (\ref{sixthh})
of the RG equation (\ref{secondd}) together with  (\ref{firstt}) and the
linear
approximation $\phi(x) = \lambda x$ can be rewritten as
\be
{d F(x) \over d\log x} = \alpha F(x)  ,
\label{ertf}
\ee
stating simply that a power law is nothing but a linear relationship when
expressed in the variables $\log F(x)$ and $\log x$. A critical point is
characterized by observables which have an invariant description with
respect to scale transformations on $x$. We can exploit this and 
the expression (\ref{ertf}) to propose the structure
of the leading corrections to the power law with log-periodicity. Hence, we
notice
that (\ref{ertf}) can be interpreted as a bifurcation equation for the
variable $F$
as a function of a fictitious ``time'' ($\log x$) as a function of the
``control
parameter'' $\alpha$. When $\alpha > 0$, $F(x)$ increases with $\log x$
while it
decreases for $\alpha < 0$. The special value $\alpha = 0$ separating the
two regimes
corresponds to a bifurcation \cite{Berge}. Once we have recognized the
structure of the expression (\ref{ertf}) in terms of a bifurcation, we can
use the general reduction theorem \cite{Berge} telling us that the
structure of
the equation for $F$ close to the bifurcation can only take a universal
{\it
non-linear} form given symmetries. Introducing the amplitude $B$ and phase
$\psi$
of $F(x)= B e^{i\psi(x)}$, the only symmetry we can use
is the fact that a global shift of the phase should keep the observable
constant under a global change of units. This implies the following
expansion\,:
\be
{dF(x) \over d\log x} =  (\alpha + i \omega) F(x)
+ (\eta + i \kappa) |F(x)|^2 F(x) + {\cal O}(F^5)   .
\label{azepo}
\ee
where $\alpha>0$, $\omega$, $\eta$ and $\kappa$ are real coefficients and
${\cal O}(F^3)$ means that higher order terms are neglected. Such
expansions
are known in the physics literature as Landau expansions
\cite{Goldenfeld}. We stress that this expression represents a non-trivial
addition to the theory, constrained uniquely by symmetry laws.

The interesting situation is the one where $x=0$ corresponds to a
critical or singular point (characterized by an unstable fixed point)  and
the other
fixed point (appearing due to the nonlinear correction) is stable. This
occurs for ${\eta \over \alpha} < 0$.
Then, small $x$ corresponds to being close to the critical point, whereas
large $x$ corresponds to the stable regime.
In terms of the amplitude $B$ and phase $\psi$ of
$F(x)= B e^{i\psi(x)}$, (\ref{azepo}) yields
\be
{\partial B \over \partial \log x} = \alpha B + \eta B^3 +... ,
\ee
\be
{\partial \psi \over \partial \log x} = \omega + \kappa B^2 +...,
\ee
whose solution reads
\be
B^2  = B_{\infty}^2 {({x \over x_0})^{2 \alpha}
\over {1 + ({x \over x_0})^{2\alpha}}} ,
\label{soklj}
\ee
\be
\psi  = \omega \log {x \over x_0} +
B_{\infty}^2 {\kappa \over 2 \alpha} \log \biggl( 1
+ ({x \over x_0})^{2\alpha} \biggl)  ,
\label{poiin}
\ee
where $B_{\infty}^2 = {\alpha \over |\eta|}$ and $x_0$ is an arbitrary
factor, fixing the time scale {\footnote{These equations correspond to
$F(x)$
describing a spiral in the complex plane, while physical values occur only
at intersections of the spiral with the real axis. This is obvious in the
linear approximation and remains true qualitatively in presence of the
non-linear
corrections. Thus, this suggests to view the discrete renormalization group
equation studied above as analogous to
discrete maps obtained from Poincar\'e sections \cite{Berge} of a
continuous renormalization group flow.}}.

Two regimes can be identified :
\begin{enumerate}
\item for small $x$ close to the singular point, the solution is of the
form
\be
B = B_{\infty} ({x \over x_0})^{\alpha} ,
\label{retgfh}
\ee
\be
\psi = \omega \log {x \over x_0}  .
\label{rdghb}
\ee
The log-periodic oscillations have a frequency equal to ${\omega \over
2\pi}$ in $\log x$.

\item For large $x$, far from the singular point, the interesting
situation is that of a stable phase at large $x$ (${\eta \over \alpha} <
0$)
and we get
\be
B \rightarrow B_{\infty} ,
\ee
\be
\psi = (\omega + B_{\infty}^2 \kappa) \log {x \over x_0}.
\ee
The log-periodic oscillations have a frequency equal to
${\omega + B_{\infty}^2 \kappa \over 2\pi}$ in $\log x$.

\end{enumerate}

For $\kappa > 0$, log-periodic oscillations, which can be detected far from
the singular point, must have a larger frequency than close to the critical
point. When going from outside the critical regime to inside the critical
regime, the frequency decreases. The reverse holds for $\kappa < 0$.

The general form of (\ref{soklj}) and (\ref{poiin}) of the solutions of the
nonlinear renormalization group equation (\ref{azepo}) lead to the
following
modification of (\ref{sixthh})\,:
\be
I\lp \tau \rp  = A + B {\lp \tau_c-\tau \rp ^{\alpha} \over \sqrt{ 1 +
\lp {\tau_c-\tau \over \Delta t} \rp^{2 \alpha}}}
\left[ 1+ C \cos \biggl( \omega \log \lp \tau_c-\tau \rp
+ {\Delta \omega \over 2 \alpha} \log \lp 1 + \lp {\tau_c-\tau \over \Delta
t} \rp^{2\alpha} \rp \biggl) \right] ,
\label{sixthhhh}
\ee
where $\tau = t/\phi$ and $\Delta \omega = B_{\infty}^2 \kappa$.

Two new effects are predicted by (\ref{sixthhhh}).
\begin{itemize}
\item There is a saturation of the function $I(\tau)$ far from the critical
point\,; 
\item the log-frequency shifts from ${\omega +
\Delta \omega \over 2\pi}$ to ${\omega \over 2\pi}$, when approaching the
time
of the crash. \end{itemize}
An interesting observation is that both effects are linked and controlled
by the same parameter $\Delta t = x_0$, which measures the characteristic
time scale controlling both the saturation and the log-frequency cross
over.

\section{Analysis of the 1929 and 1987 crashes}

We have used (\ref{sixthhhh}) to fit the Dow Jones index prior to the 1929
crash and the $S\&P500$ index prior to the 1987 crash, both starting 
approximately $8$ years prior to the crash. 

It is not clear a priori what is a good measure of the ``state'' of the
market.
In \cite{Sornette}, we used the simplest and most straightforward
parameter,
namely the market index itself. Here, we take a slightly more
sophisticated approach and test for other possibilities. The logarithm of
the index
can be argued to be a better choice. The reason is that the average growth
of the
index over the century is well-captured by an exponential rise with a
typical rate of
about $7 \%$ per year. Over $8$ years, this gives an increase by a factor
$\approx
1.7$, which is not negligible. This long-term average exponential growth is
not the
phenomenon that we are trying to detect as a signature of cooperative
behavior but
rather reflects the global price index variation as well as the global
economic
behavior. Instead of detrending the index by an average exponential growth
with the
caveat that it might distort the signal, we propose to use the logarithm of
the index
as a {\it non-parametric} proxy for the possible cooperative market
behavior. Notice
that our previous fit over 2 years has been done directly on the index as
the exponential drift has only a minor influence over this restricted time
scale.
Pushing this line of reasoning further, we have also tested whether the
subtraction
of the average $7 \%$ yearly growth to the logarithm of the index modified
the
results. We now present the results obtained for the index $I$, for $\ln I$
and for 
$\ln I -  r \tau$, where $r$ is the average yearly growth rate.

The fitting was done as a minimization of the  root-mean-square (r.m.s.) as
cost-function, assuming gaussian distributed  fluctuations, and using
Downhill
Simplex as minimization algorithm \cite{nume}. Since we are
fitting a highly non-linear function with a priori $9$  parameters to noisy
data,
many local minima exist for the cost-function. Hence, the fitting was done
in a
rather elaborate way. First, the effective number of parameters was reduced
by minimizing the cost-function with respect to the three linear variables 
$A,B,C$, thus determining $A,B,C$ explicitly as a function of the six
nonlinear variables $\alpha,t_c, \Delta t, \omega, \Delta \omega$ and
$\phi$.
This procedure thus reduces the number of free variables in the fit from
$9$ to $6$.
Due to  the risk of the minimization algorithm getting trapped in one of
the many
potential local minima, a preliminary scan, or a so-called Taboo search
\cite{taboo},
was made using a range of physically reasonable values for  $t_c, \Delta t,
\omega$
and $\Delta \omega$ and fitting only $\alpha$ and  $\phi$. To be specific,
this means that $\Delta t$ cannot have a value that is much larger or much 
smaller than the time interval of the data, since it measures the
characteristic time scale controlling both the saturation and the
log-frequency cross over. Also, very large values for $\omega$ cannot be 
accepted 
either, since it means that we are fitting fluctuations on very short  time
scales,
{\it i.e.}, ``noise''.

After the scan, all minima satisfying $0 < \alpha < 1$ was selected and are
taken as
the starting values of fits with (\ref{sixthhhh}) to the data with all $6$
non-linear
variables free. However, since  $\phi$ is just a time unit and only depends
on
whether we count in days, months or years, the fit is essentially
controlled by the
five parameters $\alpha,  t_c, \Delta t, \omega, \Delta \omega$ and only
those minima
with financially  reasonable parameter values for those parameters have
been taken
into account. For the 1929 crash, this means that between the $3$ solutions
with similar r.m.s.  (within $\approx 0.5$ \%) the solution
having $\Delta
t$ closest to the  time-interval of the date has been chosen as the best
fit.

In figures \ref{bestfit87}a and \ref{bestfit29}a, the best fits to the
logarithm of the $S\&P500$ index prior to the 1987 crash and to the
logarithm of the 
Dow Jones index prior to the 1929 crash are shown. We see that the general
trend of
the data is well-captured by the proposed relation over {\em more than 7
years}. In
order to quantify this statement, the relative error of the fit to the data
has been
calculated, see figures \ref{bestfit87}b and \ref{bestfit29}b, and the
error is $\stackrel{<}{\sim} 10 \%$ on the entire time interval.  In figure
\ref{bestfit87}a, the
thin line represents the best fit with equation (\ref{sixthhhh}) over the
whole time
interval, while the thick line is the fit by (\ref{sixthh}) on the
subinterval from
July $1985$ to the end of $1987$ as done in \cite{Sornette} but is
represented on the
full time interval starting in $1980$. The comparison with the thin line
allows one to
visualize the frequency shift described by (\ref{sixthhhh}).

Using the indexes $I$ themselves gave similar solutions, but the fitting
were 
quite unstable and with a lot of degeneracy. We attribute this to the 
superposition of the average exponential trend. Also, the fits of 
$\ln I -  r t$ gave results very similar to those shown in figures 1 and 2.

In general, the time of crash $\tau_c$ changes very little ($ < 2 \%$) 
and the log-frequency $\omega$ only moves by $\approx 25 \%$ for the best
fits 
using the three different measures. This is to be expected as the time
scale is
not modified. However, the modulation of the log-frequency, determined by
$\Delta
t$ and  $\Delta \omega$, is the less constrained than $\tau_c$ and $\omega$
and 
changes significantly as we go from $I$ to $\ln I$ or to $\ln I - r \tau$.
Furthermore, the exponent $\alpha$ increases (by a factor of 
$\approx 3$) for the best fits as the $I$-axis is compressed by 
the detrending (when going from $I$ to $\ln I$).

Observe that on relatively short time scales, we see many jumps not
accounted for by
(\ref{sixthhhh}) and obviously other processes than the ones considered
here are
influencing the stock market behavior on these time scales. Also, higher
order terms
of the solution of the renormalization group equation are not included  in
(\ref{sixthhhh}). They could play an important role at shorter time scales
as the
complete solution to all orders is expected to lead to a ``fractal''
structure with
wavy patterns at all scales.

\section{Discussion}

To validate the proposed model requires that one obtains a good fit in
several data
sets with approximately the same parameter values. Stated somewhat
pointedly: one
good fit make a data description; two good fits make a system description.
In both
cases, the fit has an overall error of less then 10 percent over a time
interval
extending up to 8 years before the crashes. There are presumably other
sorts of fits
that would work, although  the fact that 8 years of data can be adjusted
with an
error of less then $10$ percent using only $5$ parameters (not counting the
arbitrary
time unit $\phi$) is rather remarkable. It is surely irrational to infer
the validity
of a description based on a single fit. If it works for many fits, however,
and if
there is a reasonable theory for it, it should have some truth in it. In
order to
qualify such a statement, we observe that the value of the exponent
$\alpha$ and the
log-frequency $\omega$ for the two great crashes are quite close to each
other. We
find $\alpha_{1929} = 0.63$ and  $\alpha_{1987} = 0.68$. This is in
agreement with
the universality of the exponent $\alpha$ predicted from the
renormalization group
theory. A similar universality is also expected for the log-frequency,
albeit with a
weaker strength as it has been shown \cite{Saleur} that fluctuations
and noise will modify $\omega$ differently depending on their nature. We
find from
the fits that $\omega_{1929} = 5.0$ and $\omega_{1987} = 8.9$. These values
are not
unexpected and correspond to what has been found for other crashes as well
as for
earthquakes \cite{SSam,Varnes} and for related
rupture and growth phenomena \cite{Anifrani,So,Salal1,Salal2}.  If we were
fitting
random fluctuations around some average power law rise, then the obtained
value of
$\omega$ would generally be higher and fluctuate more due to the
noise-fitting we
would be performing. Furthermore, the values obtained for the amplitude $C$
shows
that a log-periodic correction to a pure power law is not insignificant.
The analysis
of the two great crashes in this century presented here  and supplementing 
\cite{SSam,Anifrani,Joh,Sornette,So} suggests a very coherent picture,
namely that
{\it complex} critical exponents are a general phenomenon in irreversible
 self-organizing systems. It is interesting that the similarity
between the
two situations in 1929 and 1987 has in fact been noticed qualitatively in
an article
in the {\it Wall Street Journal} on october 19, 1987, the very morning of
the day of
the stock market crash (with a plot of stock prices in the 1920s and the
1980s). See
the discussion in \cite{Shiller}.

The stock market provides a remarkable realization
of a complex self-organizing system and the
log-periodic structure found prior to crashes implies the existence of a
hierarchy of characteristic time scales, corresponding to the time
intervals $t_c - t_n$, determined from the equation
$\omega \log (t_c-t_n) + \phi  =  n 2 \pi$ for
which the cosine in (\ref{sixthhhh}) is largest. These time scales could
reflect the characteristic relaxation times associated with the coupling
between traders and the fundamentals of the economy. The larger value of
$\omega$ (smaller ratio $\lambda$) for the more recent crash could reflect
the faster nature of the fluctuations resulting from more efficient
computerized trading systems. While $\alpha$ is expected to remain robust, future
crashes should have a similar or even larger value $\omega$.

The main point of this paper is that the market anticipates the crash in a
subtle self-organized and cooperative fashion, hence releasing precursory
``fingerprints'' observable in the stock market prices. In other words,
this implies that market prices contain information on impending crashes.
If
the traders were to learn how to decipher and use this information, they
would act on it and on the knowledge that others act on it and the crashes
would
probably not happen. Our results suggest a weaker form of the ``weak
efficient
market hypothesis'' \cite{Fama}, according to which the market prices
contain,
in addition to the information generally available to all, subtle
informations
formed by the global market that most or all individual traders have not
yet learned to decipher and use. Instead of the usual interpretation of the
efficient market hypothesis in which traders extract and incorporate
consciously (by their action) all informations contained in the market
prices, we propose that the market as a whole can exhibit an ``emergent''
behavior
not shared by any of its constituent. In other words, we have in mind the
process of the emergence of intelligent behavior at a macroscopic scale
that
individuals at the  microscopic scale have not idea of. This process has
been
discussed in biology for instance in animal populations such as ant
colonies or in
connection with the emergence of consciousness \cite{Anderson,Holland}. The
usual
efficient market hypothesis will be recovered in this context when the
traders learn
how to extract this novel collective information and act on it.

Most previous models proposed for crashes have pondered the possible
mechanisms to
explain the collapse of the price at very short time scales. Here in
contrast, we
propose that the underlying cause of the crash must be searched years
before it in
the progressive accelerating ascent of the market price, reflecting an
increasing
built-up of the market cooperativity. From that point of view, the specific
manner by
which prices collapsed is of not of real importance since, according to the
concept
of the critical point, any small disturbance or process may have triggered
the
instability. The intrinsic divergence of the sensitivity and the growing
instability
of the market close to a critical point might explain why attempts to
unravel the
local origin of the crash have been so diverse. Essentially, anything would
work once
the system is ripe. Our view is that the crash has an endogenous origin and
that
exogenous shocks only serve as triggering factors. We propose that the
origin of 
crashes is much more subtle and is constructed progressively by the market
as a whole.
In this sense, this could be termed a systemic instability.

Acknowledgements : We thank M. Brennan, O. Ledoit, W.I.
Newman and H. Saleur for useful discussions. A. Johansen thanks the SARC
Foundation for support.

\pagebreak

\pagebreak

{\Large \bf Figure captions}

\newcounter{fig}
\usecounter{fig}
\begin{list}{Figure: \arabic{fig}}
{\usecounter{fig}\setlength{\labelwidth}{2cm}\setlength{\labelsep}{3mm}}

\item \label{bestfit87}  a) Time dependence of the logarithm of the New
York stock
exchange index $S\&P500$ from january 1980 to september 1987
and best fit by (\ref{sixthhhh}) (thin line). The crash of October 14, 1987
corresponds  to $1987.78$ decimal years. The thin line represents the best
fit with equation  parameters of the fit are: r.m.s.$=0.043$, $t_c =
1987.81$
year, $\alpha = 0.68$, $\omega = 8.9$,   $\Delta \omega = 18$, $\Delta t =
11$
years, $A = 5.9$, $B = -0.38$,  $C = 0.043 $. The thick line is the fit by
(\ref{sixthh}) on the subinterval from July $1985$ to the end of $1987$ and
is represented on the full time interval starting in $1980$. The parameters
of this fit with (\ref{sixthh}) are r.m.s=$6.2$, $t_c = 1987.74$ year,
$\alpha 
= 0.33$, $\omega = 7.4$, $A =  412$,  $B=  -165$, $C= 0.07$. The comparison
with the thin line allows one to visualize the frequency shift described by
(\ref{sixthhhh}).

b) The relative error of the fit by (\ref{sixthhhh}) to the data.

\item \label{bestfit29}  a) Time dependence of the logarithm of the Dow
Jones stock
exchange index from june 1921 to september 1929 and best fit by 
(\ref{sixthhhh}). The crash of October 23, 1929 corresponds to $1929.81$ 
decimal years. The parameters of the fit are:  
r.m.s.$=0.041$, $t_c = 1929.84$ year, $\alpha = 0.63$, $\omega = 5.0$,
$\Delta \omega = -70$, $\Delta t = 14$ years, $A = 61$, $B = -0.56$,
$C = 0.08$.

b) The relative error of the fit by (\ref{sixthhhh}) to the data.

\end{list}

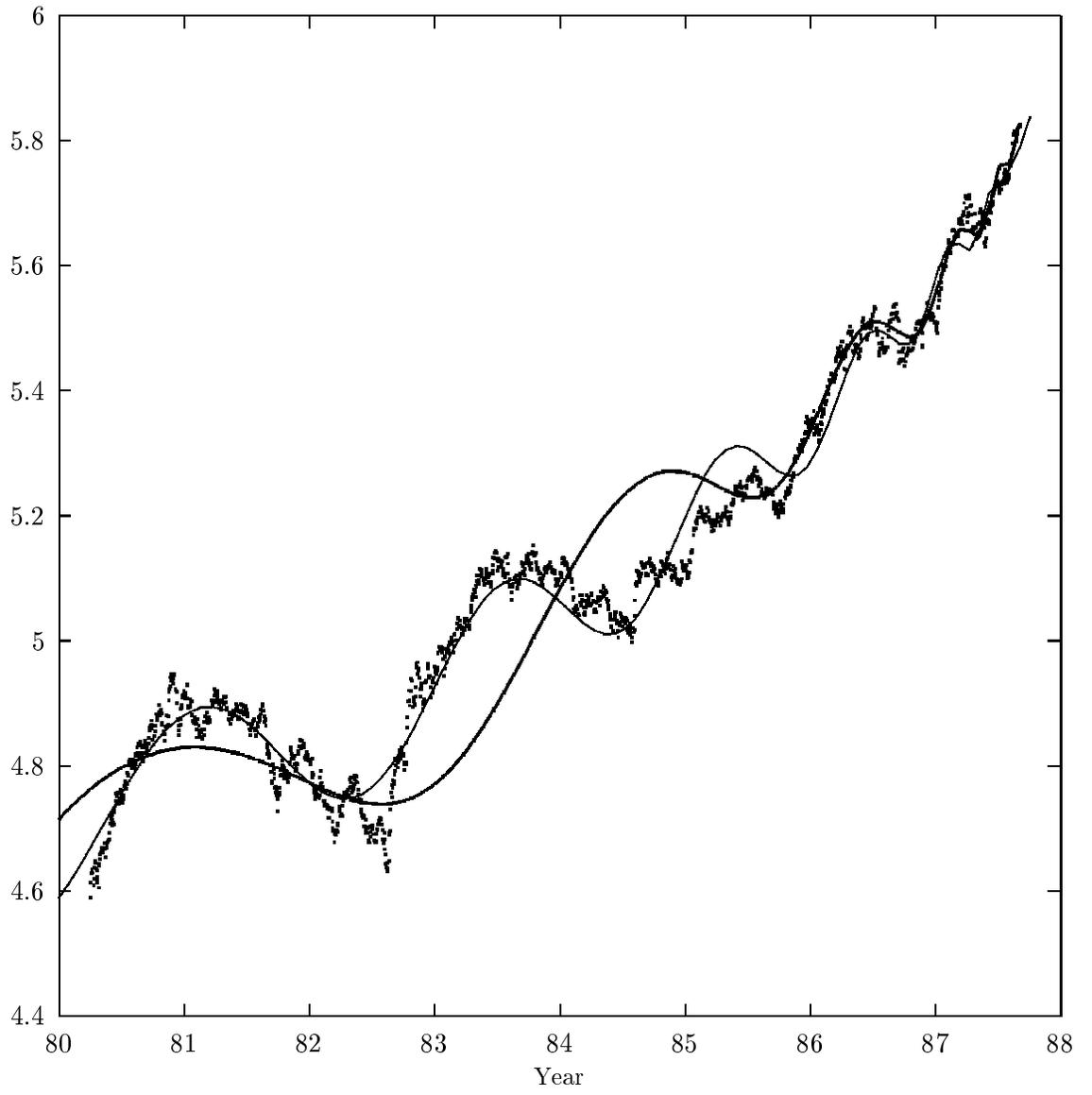
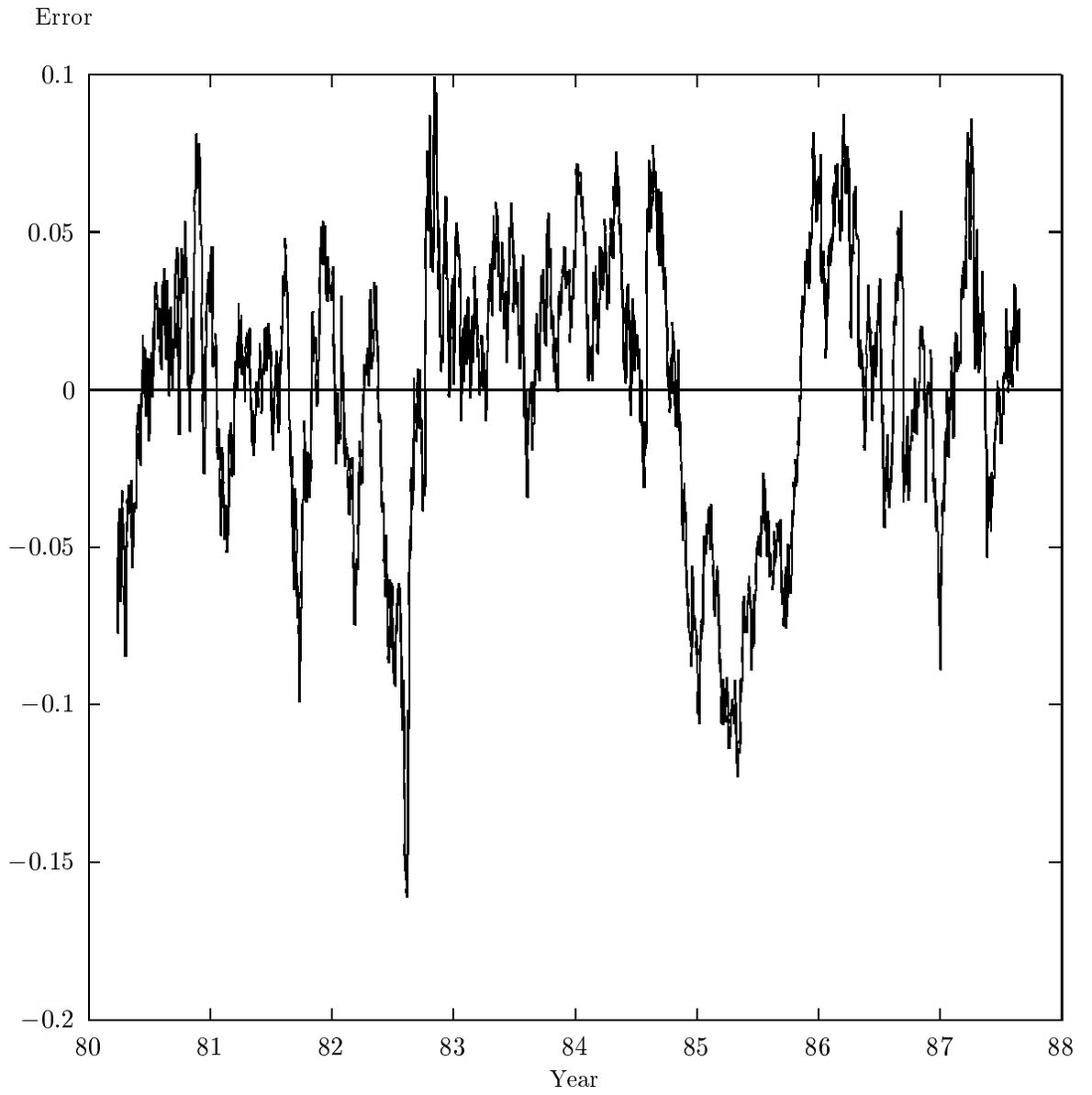
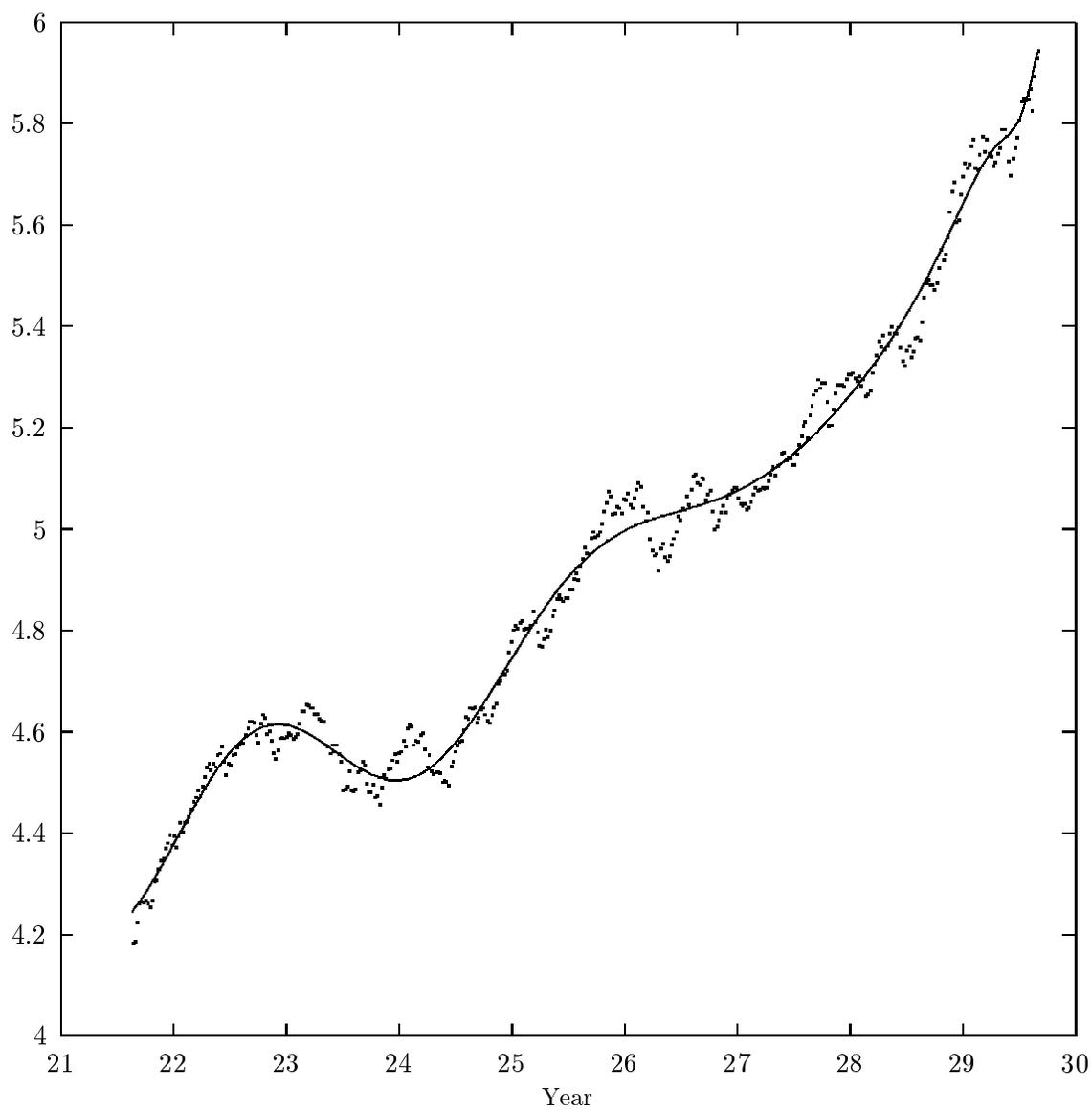
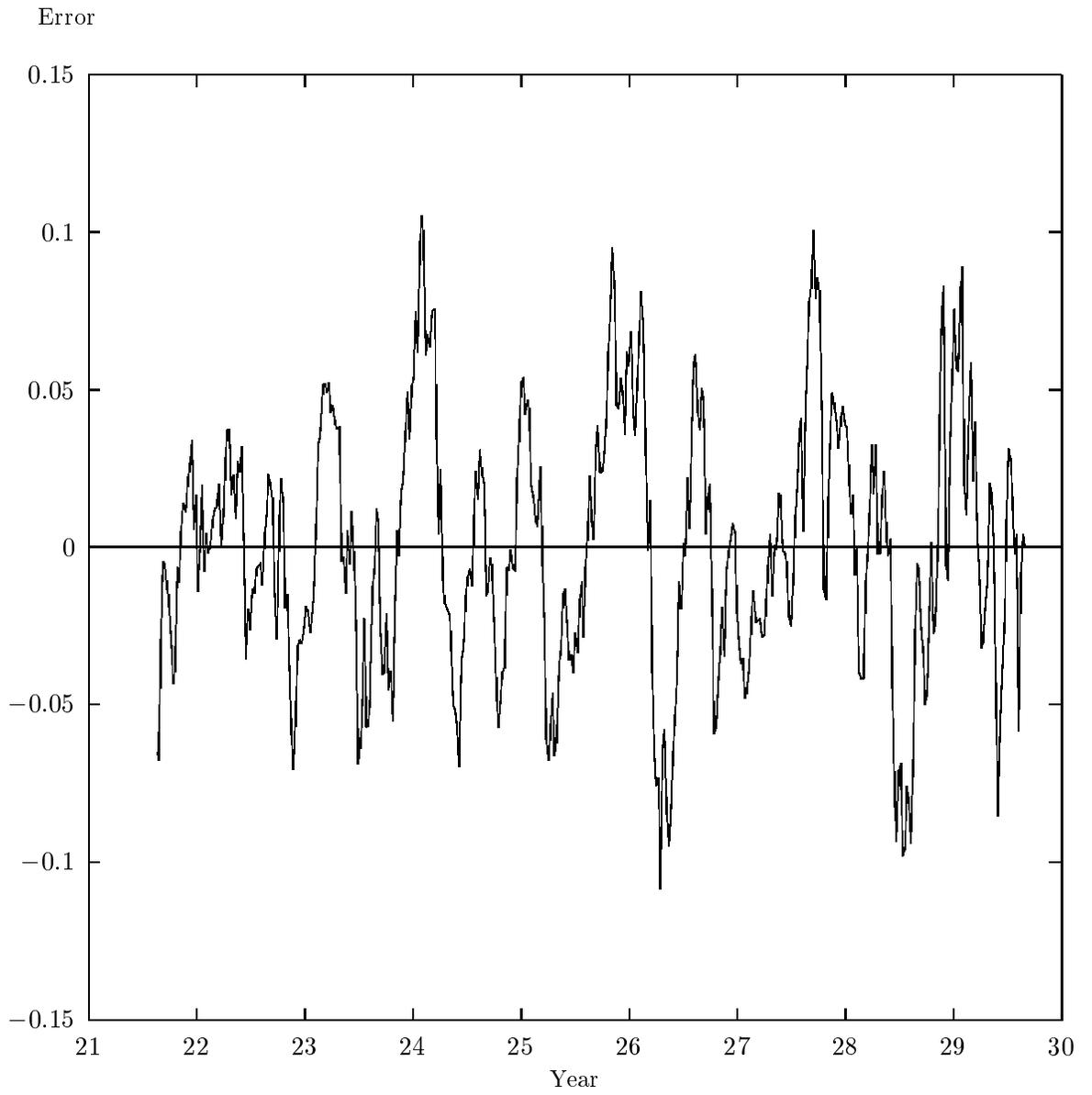
\end{document}